# Surface Integral Formulation with Preconditioned, Multipole-Accelerated Iterative Method Applied for the Analysis of RCS of Single and Multi-Layer Dielectric Structures

Hamid T. Chorsi
University of Colorado, Denver, USA

**ABSTRACT**

A preconditioned, multipole-accelerated, Krylov-subspace iterative algorithm for the electromagnetic scattering analysis of three dimensional (3D), arbitrary shaped dielectric structures composed of single and multi-layered dielectric materials is presented. A surface integral equation (SIE) formulation using Maxwell's equations and the equivalence principle which produces a well-conditioned matrix equation is applied. The surface integral equation is discretized using the well-known Method of Moments (MoM) based on a Galerkin scheme. The dense matrix is solved efficiently via an iterative method accelerated by the Multilevel Fast Multipole Method (MLFMM) coupled with an incomplete LU (ILU) preconditioning technique. Compared to the proposed methods in the literature, proposed method is much simpler and efficient to implement for layered dielectric structures. 3D numerical examples are presented to demonstrate the validity and accuracy of the proposed approach.

## Introduction

Single and multi-layer dielectric structures used for waveguides, antennas, filters, nanoantennas, etc. have received significant attention during the last two decades because of their advantages over conventional metallic structures which suffer from problems like power loss, low bandwidth, low radiation efficiency and fabrication difficulties [1]. Figure 1 shows two simple configurations of multi-layer dielectric antennas which comprises different dielectric materials with different permittivity.

There is a clear need for computational and numerical analysis and design tools for calculating the performance and optimizing the parameters of dielectric structures prior to costly prototype development. These computational methods are the building blocks of the commercial simulation software for modeling of three-dimensional metallic, metallic-dielectric, and all-dielectric structures. One of the most general tactics in analysis of dielectric structures is the finite-difference time-domain (FDTD) method. FDTD has been used by a number of authors to analyze dielectric structures [2-5] [6]. Although the FDTD is capable of modeling and analyzing an extensive range of arbitrary-shaped dielectric structures, the computational cost is very problem dependent and it suffers from the stair-casing problems [7].

The surface integral equation method is one of the most popular numerical methods in electromagnetic analysis of homogenous dielectric and composite metallic and dielectric objects. The SIE has been extensively used in the literature for the analysis of different electromagnetic problems. In [8] SIE based on the method of moments (MoM) has been used to simulate the interaction of light and plasmonic nanostructures. SIEs are also popular in the microwave regime, and have been extensively used for the analysis of periodic lossy [9] or metallic [10] structures. Recently, many efforts have been made to improve the efficiency of the dielectric formulations. In [11] authors have proposed a new surface integral equation formulation namely the combined tangential formulation (CTF) for the analysis of dielectric structures. Despite its accuracy, it requires a large number of iterations which limits the practicality of the formulation for the real life complex applications.

In this paper, a surface integral equation (SIE) formulation to analyze electromagnetic scattering and radiation problems of composite dielectric structures is derived. The SIE is discretized via the well-known method of moments (MoM) using a Galerkin procedure. The resulting dense matrix is solved via a preconditioned iterative solution accelerated by the multilevel fast multipole method (MLFMM) to speed up vector-matrix multiplications. A complete analysis of the SIE formulations proposed in the literature with different iterative solvers and preconditioner is conducted in the paper in order to find their impact on the scattering analysis of layer structures. Several examples at different frequencies and for different sizes is also presented to show the flexibility and applicability of the proposed method. It has been shown that the proposed SIE can be extremely useful for rapidly prototyping of novel layered structures.

## Integral Equation Formulation

Consider the case in Figure 2 involving the problem of electromagnetic scattering by an arbitrary shaped, homogenous dielectric object $B_2$ characterized by electrical properties $(\varepsilon_2, \mu_2)$ which is surrounded by a homogeneous (unbounded) medium $B_1$ having electrical properties $(\varepsilon_1, \mu_1)$.

Let $\Gamma_{1,2}$ be the surface separating media $B_1$ and $B_2$. Also, let $\Gamma_{1,2}^l$ represent the surface just inside medium $B_l$, where $l = 1, 2$. Equivalent current densities can then be introduced on $\Gamma_{1,2}^l$ as:

$$\hat{n}_l \times E_l^{inc} + \hat{n}_l \times E_l^{scat} = -M_l \tag{1}$$

$$\hat{n}_l \times H_l^{inc} + \hat{n}_l \times H_l^{scat} = J_l \tag{2}$$

where, $\hat{n}_l$ is the *inward* normal of $B_l$ on $\Gamma_{1,2}^l$, as illustrated in Figure 2, $(E_l^{inc}, H_l^{inc})$ are the incident fields radiated from within $B_l$, and $(E_l^{scat}, H_l^{scat})$ are the scattered fields radiated by the equivalent currents $(J_l, M_l)$ on $\Gamma_{1,2}^l$ inside of an effective unbounded medium $(\varepsilon_l, \mu_l)$, where

$$E_l^{scat} = \eta_l \mathcal{L}_l \{J_l\} - \mathcal{K}_l \{M_l\}_l \tag{3}$$

$$H_l^{scat} = \mathcal{K}_l \{J_l\} + \eta_l^{-1} \mathcal{L}_l \{M_l\} \tag{4}$$

The operators $\mathcal{L}$ and $\mathcal{K}$ are defined as:

$$\mathcal{L}_l \{X\} = jk_l[\int_s X(r')ds' + k_l^{-2} \int_s ds' \nabla' \cdot X(r') \nabla] G_l(r, r') \tag{5}$$

$$\mathcal{K}_l \{X\} = \int_s X(r') \times \nabla' G_l(r, r') ds' \tag{6}$$

Where $k_l = \omega \sqrt{\mu_l \varepsilon_l}$ is the wavenumber, $\eta_l = \sqrt{\mu_l / \varepsilon_l}$ is the characteristic wave impedance, $\nabla' \cdot$ denotes the divergence in the primed (source) coordinates and

$$G_l(r, r') = \frac{e^{-jk_l R}}{4\pi R} \qquad (R = |r - r'|) \tag{7}$$

is the homogenous Green's function.

Combining (1) – (6), leads to

N-EFIE(l)  $\qquad \eta_l \hat{n}_l \times \mathcal{L}_l \{J_l\} + M_l - \hat{n}_l \times \mathcal{K}_l \{M_l\} = -\hat{n}_l \times E_l^{inc} \tag{8}$

N-MFIE(l)  $\quad -\mathbf{J}_l + \hat{n}_l \times \boldsymbol{K}_l\{\mathbf{J}_l\} + \eta_l^{-1}\hat{n}_l \times \mathcal{L}_l\{\mathbf{M}_l\} = -\hat{n}_l \times \mathbf{H}_l^{inc}$ (9)

which will be referred to as the Normal-EFIE (N-EFIE) and Normal-MFIE (N-MFIE)[1], respectively, for medium $B_l$. The tangential equations are the derived directly from (8) and (9) as

T-EFIE(l) $\quad -\hat{n}_l \times \mathbf{M}_l + \hat{n}_l \times \hat{n}_l \times \boldsymbol{K}_l\{\mathbf{M}_l\} - \eta_l \hat{n}_l \times \hat{n}_l \times \mathcal{L}_l\{\mathbf{J}_l\} = \hat{n}_l \times \hat{n}_l \times \mathbf{E}_l^{inc}$ (10)

T-MFIE(l) $\quad \hat{n}_l \times \mathbf{J}_l - \hat{n}_l \times \hat{n}_l \times \boldsymbol{K}_l\{\mathbf{J}_l\} - \eta_l^{-1}\hat{n}_l \times \hat{n}_l \times \mathcal{L}_l\{\mathbf{M}_l\} = \hat{n}_l \times \hat{n}_l \times \mathbf{H}_l^{inc}$ (11)

The equivalent currents $\vec{\mathbf{J}}_l$ and $\vec{\mathbf{M}}_l$ on either side of the boundary $\Gamma_{1,2}^l$ can be related by enforcing the tangential boundary conditions:

$$\mathbf{J}_1(r') = \hat{n}_1 \times \mathbf{H}_1(r') = -\hat{n}_2 \times \mathbf{H}_2(r') = -\mathbf{J}_2(r') \quad (12)$$

$$-\mathbf{M}_1(r') = \hat{n}_1 \times \mathbf{E}_1(r') = -\hat{n}_2 \times \mathbf{E}_2(r') = \mathbf{M}_2(r') \quad (13)$$

where, $n_1$ and $n_2$ represent the inward unit normal on dielectric 1 and 2, respectively as again shown in Figure 3.

Among the various possibilities of combinations of the normal and tangential electric and magnetic fields, a mixed formulation called the electric and magnetic current combined field integral equation (JMCFIE) [12-15], is used, which is expressed as:

$$\begin{bmatrix} \sum_{l=1}^{2}\eta_l^{-1}\{\text{T-EFIE(l)}\} + \sum_{l=1}^{2}(-1)^l\{\text{N-MFIE(l)}\} \\ \sum_{l=1}^{2}\eta_l\{\text{T-MFIE(l)}\} - \sum_{l=1}^{2}(-1)^l\{\text{N-EFIE(l)}\} \end{bmatrix} \quad (14)$$

Applying the boundary conditions of (12) and (13), (14) can be written as:

$$\begin{bmatrix} [Z_{11}] & [Z_{12}] \\ [Z_{21}] & [Z_{22}] \end{bmatrix} \cdot \begin{bmatrix} \mathbf{J} \\ \mathbf{M} \end{bmatrix} = \begin{bmatrix} \sum_{l=1}^{2}\left(\eta_l^{-1}(\mathbf{E}^{inc})_{\tan} - \hat{n}_l \times \mathbf{H}^{inc}\right) \\ \sum_{l=1}^{2}\left(\eta_l(\mathbf{H}^{inc})_{\tan} + \hat{n}_l \times \mathbf{E}^{inc}\right) \end{bmatrix} \quad (15)$$

---

[1] While the N-EFIE and N-MFIE represent tangential projections of the electric and magnetic fields, respectively, the operator is based on $\hat{n}_l \times$ which rotates the tangential field about the normal by 90-degrees.

For a composite dielectric object involving $N_\Gamma$ boundaries $\Gamma_{l,k}$ separating medium $B_l$, and $B_k$ with constitutive parameters $(\varepsilon_l, \mu_l)$ located in a host medium $(\varepsilon_k, \mu_k)$, the surface integral equation can be derived as

$$\begin{bmatrix} [\sum_{k=1}^{N_\Gamma} Z_{11,k}] & [\sum_{k=1}^{N_\Gamma} Z_{12,k}] \\ [\sum_{k=1}^{N_\Gamma} Z_{21,k}] & [\sum_{k=1}^{N_\Gamma} Z_{22,k}] \end{bmatrix} \cdot \begin{bmatrix} J \\ M \end{bmatrix} = \begin{bmatrix} \sum_{p=1}^{N_\Gamma} \sum_{l=1}^{2} (\eta_l^{-1}(E^{inc})_{tan} - (-1)^l \hat{n}_l \times H^{inc})_{1,p} \\ \sum_{p=1}^{N_\Gamma} \sum_{l=1}^{2} (\eta_l (H^{inc})_{tan} + (-1)^l \hat{n}_l \times E^{inc})_{2,p} \end{bmatrix} \quad (16)$$

## Discretization and Iterative Solution

To obtain the equivalent electric (J) and magnetic (M) currents on the surface of the dielectric object, the well-known method of moments (MoM) is applied to the integral equation in (12). The equivalent currents are expanded into a series of known basis functions $b_n(r)$ (n=1, 2,…, N) on the surface S in the following form

$$J(r) = \sum_{n=1}^{N} c_n b_n(r) \qquad r \in S \quad (13)$$

$$M(r) = \sum_{n=1}^{N} v_n b_n(r) \qquad r \in S \quad (14)$$

Where $c_n$ and $v_n$ are the unknown expansion complex coefficients. Substituting equations (13) and (14) into equation (12) and applying the Galerkin scheme (same basis and testing functions $t_m(r)$), the following linear equation can be derived:

$$\begin{bmatrix} \ell_l^T + \ell_l^N & \frac{1}{\eta_l}\left[\kappa_l^N - \kappa_l^T\right] \\ \eta_l\left[\kappa_l^T - \kappa_l^N\right] & \ell_l^T + \ell_l^N \end{bmatrix} \cdot \begin{bmatrix} J \\ M \end{bmatrix} = \begin{bmatrix} E_l \\ H_l \end{bmatrix} \quad (15)$$

Where

$$\ell_l^T[m,n] = j\partial_m\partial_n k_l \int_{S_m} t_m(r)dr \times \int_{S_n} G(r,r') \, b_n(r')dr' + \tag{16}$$

$$j\partial_m\partial_n k_l^{-1} \int_{S_m} t_m(r)dr \times \int_{S_n} \nabla G(r,r') \, b_n(r')dr'$$

$$\ell_l^N[m,n] = j\partial_n k_l \int_{S_m} t_m(r)dr \cdot \vec{n}_m \times \int_{S_n} G(r,r') \, b_n(r')dr' + \tag{17}$$

$$j\partial_n k_l^{-1} \int_{S_m} t_m(r)dr \cdot \vec{n}_m \times \int_{S_n} \nabla G(r,r') \nabla' b_n(r')dr'$$

$$\kappa_l^N[m,n] = \partial_n \int_{S_m} t_m(r)dr \cdot \vec{n}_m \times \int_{S_n} \nabla' G(r,r') \, b_n(r')dr' \tag{18}$$

$$\kappa_l^T[m,n] = \partial_m\partial_n \int_{S_m} t_m(r)dr \cdot \int_{S_n} \nabla' G(r,r') \, b_n(r')dr' \tag{19}$$

And the elements of the right hand side vector are derived as

$$E_l = -\partial_m \frac{1}{\eta_l} \int_{S_m} t_m(r) \cdot E_l^{inc}(r)dr - \int_{S_m} t_m(r) \cdot \hat{n}_m \times H_l^{inc}(r)dr \tag{20}$$

$$H_l = -\partial_m \eta_l \int_{S_m} t_m(r) \cdot H_l^{inc}(r)dr + \int_{S_m} t_m(r) \cdot \hat{n}_m \times E_l^{inc}(r)dr \tag{21}$$

In which $\partial_n$ and $\partial_m (\pm 1)$ are the orientations of the basis and testing functions (c.f., Figure 4), and $\hat{n}_m$ is the normal vector of the testing function.

Here we employ the extensively used Rao–Wilton–Glisson (RWG) functions [16] as basis and testing functions, which discretizes the surface of the dielectric object into small elements each of which is triangle. Figure 4 depicts an RWG function on a pair of triangles.

An RWG function associated with the $i$ th edge can be defined as

$$f_i(r) = \begin{cases} \dfrac{l_i \rho_i^+}{2\Lambda_i^+}, & r \in T_i^+ \\ -\dfrac{l_i \rho_i^-}{2\Lambda_i^-}, & r \in T_i^- \\ 0, & otherwise \end{cases} \tag{22}$$

Where $l_i$ is the length of the main edge, $\Lambda_i^\pm$ is area of triangle $T_i^\pm$ associated with the edge, and $\rho_i^+$ and $\rho_i^+$ are respectively radial vectors with respect to the free vertex of $T_i^+$ and $T_i^-$. Also note that in Figure 4, the total charge is zero for each RWG function, ensuring that all edges of $T_i^+$ and $T_i^-$ are free of line charges.

Discretization using RWG functions will lead to the following operators in (15)

$$\ell_l^T[m,n] = \frac{\partial_m \partial_n jk_l l_m l_n}{4\Lambda_m \Lambda_n} \int_{S_m} \rho_m^+ dr \cdot \int_{S_n} G_l(r,r')\rho_n^- dr' - \frac{\partial_m \partial_n jk_l^{-1} l_m l_n}{\Lambda_m \Lambda_n} \int_{S_m} dr \cdot \int_{S_n} G_l(r,r')dr' \quad (23)$$

$$\ell_l^N[m,n] = \frac{\partial_n jk_l l_m l_n}{4\Lambda_m \Lambda_n} \int_{S_m} \hat{n} \times \rho_m^+ dr \cdot \int_{S_n} G_l(r,r')\rho_n^- dr' - \frac{\partial_n jk_l^{-1} l_m l_n}{2\Lambda_m \Lambda_n} \int_{S_m} \hat{n} \times \rho_m^+ dr \cdot \int_{S_n} \nabla G_l(r,r')dr' \quad (24)$$

$$\kappa_l^N[m,n] = \frac{\partial_n l_m l_n}{4\Lambda_m \Lambda_n} \int_{S_m} \hat{n} \times \rho_m^+ dr \cdot \rho_n^+ \times \int_{S_n} \nabla G_l(r,r')dr' \quad (25)$$

$$\kappa_l^T[m,n] = \frac{\partial_m \partial_n l_m l_n}{4\Lambda_m \Lambda_n} \int_{S_m} \rho_m^+ dr \cdot \rho_n^+ \times \int_{S_n} \nabla G_l(r,r')dr' \quad (26)$$

Once the current densities are computed using (15), the scattered electric field can be computed. . In the far-field scattered electric field can be calculated as:

$$E^{sca} = \frac{j\omega\mu_0}{4\pi}\left[\int_S J(r')e^{-jk_0 \cdot r'}dr' - \eta_0^{-1}\int_S M(r')e^{-jk_0 \cdot r'}dr'\right] \quad (27)$$

Using (13) and (14) and then discretizing using (22) the far electric field is computed from the method of moment solution as:

$$E^{sca}(\theta,\phi) = \frac{j\omega\mu_0}{4\pi}\left[\int_S \sum_{n=1}^N c_n b_n(r')e^{-jk_0 \cdot r'}dr' - \eta_0^{-1}\int_S \sum_{n=1}^N v_n b_n(r')e^{-jk_0 \cdot r'}dr'\right]$$
$$= \pm\frac{j\omega\mu_0}{4\pi} \cdot \sum_{n=1}^N c_n \sum_{q=1}^2 \frac{l_n}{2\Lambda_{nq}}\int_S \rho_n^+ e^{-jk_0 \cdot r'}dr' - \frac{jk_0}{4\pi} \cdot \sum_{n=1}^N v_n \sum_{q=1}^2 \frac{l_n}{2\Lambda_{nq}}\int_S \rho_n^+ e^{-jk_0 \cdot r'}dr' \quad (28)$$

In which the electric (J) and magnetic (M) currents obtain from (15). Finally, the radar cross section (RCS) is computed from the far electric field using

$$RCS = \lim_{R \to \infty} \left[ 4\pi R^2 \frac{\left|E^{sca}(\theta,\phi)\right|^2}{\left|E^{inc}(\theta,\phi)\right|^2} \right] \quad (E^{sca,inc} = E_\phi \text{ or } E_\theta) \quad (29)$$

## Simulation Results

In this section, we investigate the validity and efficiency of the proposed method for the efficient analysis of single and multilayer dielectric structures. First, we examine the accuracy of the proposed method by comparing our simulation results for a two layer dielectric sphere with the analytical solution provided by the Mie series. This verification is essential in order to guarantee that both formulation are consistent.

Figure 5 shows the bistatic RCS values for a two-layer dielectric sphere, located in free space and discretized with $\lambda/10$ mesh sizes. The normalized RCS is plotted as a function of the observation angle which is from 0° to 180°, where 0° is the forward-scattering direction. It can be observed that the computational values from JMCFIE with the Multilevel Fast Multipole Method (MLFMM) [17] are in good agreement with results from the analytical results.

Figure 6 presents four different SIE formulations for the two-layer sphere in Figure 5. It can be seen from Figure 6 that by increasing the number of unknowns (decreasing the mesh size) the number of iterations are getting much higher for the combined tangential formulation (CTF) (which involves a careful and improved scaling of T-EFIE and T-MFIE) [18] compared to our derivation and implementation of the JMCFIE. From the figure it also can be observed that the iteration number for the combined normal formulation (CNF) [18] (which is the combination of N-EFIE and N-MFIE) is better than the CTF, but as the number of unknowns increases the iteration number of the CNF is growing faster than the CTF. For the improved CTF (ICTF) [19] the number of iterations is higher than original CTF. For the well-known Poggio-Miller-Chang-Harrington-Wu-Tsai (PMCHWT) the number of iterations are much higher that CTF.

For dense-matrix formulations many Krylov-subspace iterative algorithms are available. Among them the biconjugate gradient-stabilized (BiCGStab) algorithm [20] is a good choice for well-conditioned matrix equations such as the JMCFIE. In this paper the solution has been accelerated using an incomplete LU (ILU) preconditioner [21] which is suitable for MLFMM [reference?]. Figure 7 presents the number of iterations required by the MLFMM accelerated JMCFIE with solved with different iterative methods.

From Figure 7 it is clear that the ILU preconditioned JMCFIE-MLFMM with BiCGStab iterative method requires fewer iterations as compared to JMCFIE-MLFMM with conjugate gradient squared (CGS) method [22]. It also requires fewer iterations than the BiCGStab iterative solver without a preconditioner (NoP).

To further investigate the validity of the proposed JMCFIE method two additional examples of all-dielectric resonator antennas are studied. The simulation results are compared with the commercial software FEKO [23].

Both objects are illuminated with plane wave excitation in the $\hat{k}_i$ direction

$$E^{inc}(r) = \hat{e}E_0 exp(ik_0\hat{k}_i \cdot \vec{r}) \tag{30}$$

Where $E_0$ is the amplitude of the plane wave, $\hat{e}$ is the polarization of the electric field, $\hat{k}$ is the wave direction, and $\hat{e} \perp \hat{k}$.

The first example is the simulation of a simple P-shape single-layer dielectric antenna which is made of a dielectric material with the relative permittivity of $\varepsilon_r = 24$. The dimensions of the antenna are presented in Figure 8-a. The antenna is placed in the x-y plane and illuminated by a plane wave (1 V/m) in the +z direction in free space (electric field is in +x direction). Figure 9-a presents the RCS (structural and antenna mode RCS [24]) of the P-shape antenna at 5.1 GHz which is in good agreement with FEKO software.

In Figure 9-a, the left-hand side figure is for the case in which theta is constant (90º) and phi is changing from 0º to 360º. On the other hand the right-hand side figure is for the case in which phi is constant (0º) and theta is changing from 0º to 180º.

As a second example we have considered the analysis of a concentric half-split dielectric resonator antenna (CDRA) which is developed using permittivity variation in radial direction. In this paper we have considered three CDRA which are proposed in [25]. The fabricated three-layer concentric dielectric resonator antennas is shown in Figure 8-b. The dielectric constant and radius of the three-layer antenna are $\varepsilon_{r1} = 10.2, \varepsilon_{r2} = 6.15, \varepsilon_{r3} = 2.32, r_1 = 5mm, r_2 = 10mm, r_3 = 20mm$ and $h = 11.4mm$.

Figure 9-b presents the simulation results for three-layer dielectric antenna. The object is placed in x-y plane in free space and illuminated with a unit plane wave (1 V/m) in +z direction and the electric filed is in +x direction. In the left-hand side figure, we keep theta constant (90º), on the other hand, in the right-hand side figure phi is constant (0º).

Finally, in order to show the applicability and the flexibility of the proposed method we have considered the simulation of a layered nanoparticle. The presented spherical nanoparticle consists of a spherical silver nanoparticle with radius of 230 nm and is coated with 20 nm of gold in order to increase their absorption rate [26]. The optical properties of gold and silver obtained from [27]. The nanoparticle is placed in the free space and illuminated by a plane wave (1 V/m) in the +z direction in free space (electric field is in +x direction). The RCS results at 545 THz (550 nm) for constant theta (90º) are presented in Figure 10. A good agreement between the proposed method and the Mie results can be observed in the Figure.

## Conclusion

A surface integral equation formulation obtained from the Maxwell's equations has been applied to the electromagnetic scattering and radiation problems of three-dimensional (3D), arbitrary shaped single and multi-layer dielectric structures. The obtained integral equations were discretized with the well-known method of moments (MoM), where the multilevel fast multipole method (MLFMM) has been employed to speed up matrix-vector multiplication from $O(N^3)$ to $O(N \log N)$ (N is the number of unknowns) and reduce memory requirements. Some 3D examples were presented to confirm the validity and versatility of this approach on dealing with dielectric antennas.

# Figures

Figure 1: multi-layer dielectric resonator structures

Figure 2: A homogenous region and boundary surfaces

Figure 3: Boundary conditions for two dielectrics

Figure 4: Illustration of the RWG basis function associated to the $l^{th}$ edge. $\partial_m = +1$ on $T_i^+$ and $\partial_m = -1$ on $T_i^-$

Figure 5: (a) a two-layer dielectric sphere (b) normalized bistatic RCS of a two-layer sphere

Figure 6: Comparison between two SIE formulations CTF and JMCFIE for a two-layer dielectric sphere in figure 5

Figure 7: Comparison between different iterative methods for JMCFIE for a two-layer dielectric sphere in 5

Figure 8: (a) structure of the P-shape resonating antenna (1mm mesh) (b) structure of the three-layer concentric antenna (1mm mesh)

Figure 9: Simulation results of the P-shape single-layer and three-layer concentric dielectric antennas (a) RCS of the P-shape antenna at 5.1 GHz (b) RCS of three-layer concentric antenna at 6 GHz

Figure 10: Simulation results of the presented layered nanoparticle at 545 THz (550 nm).

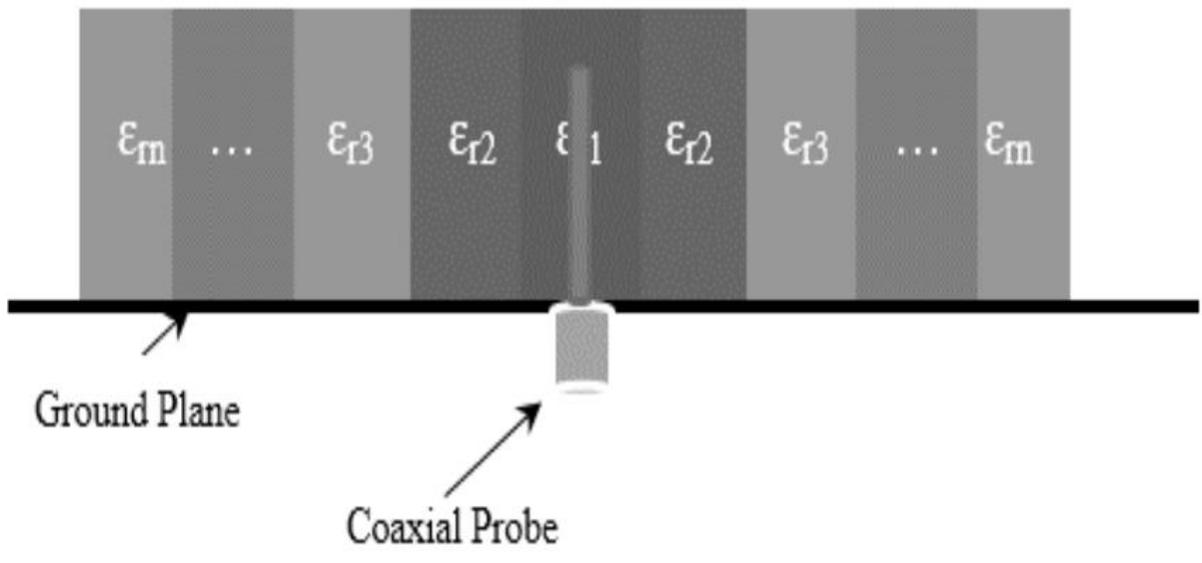

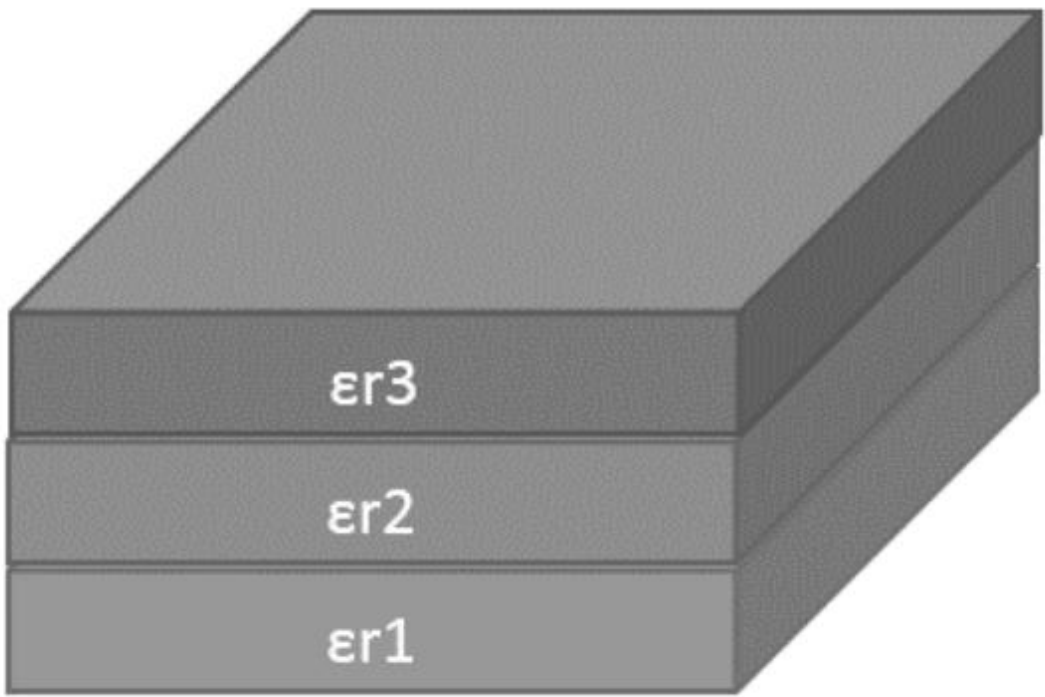

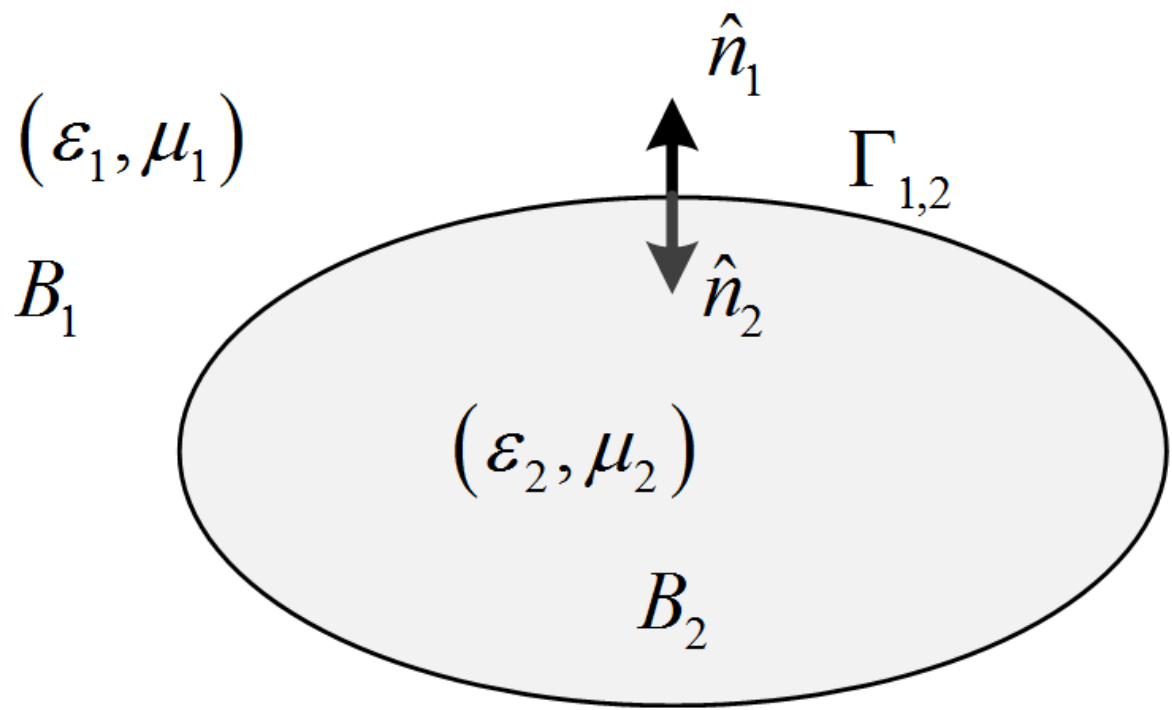

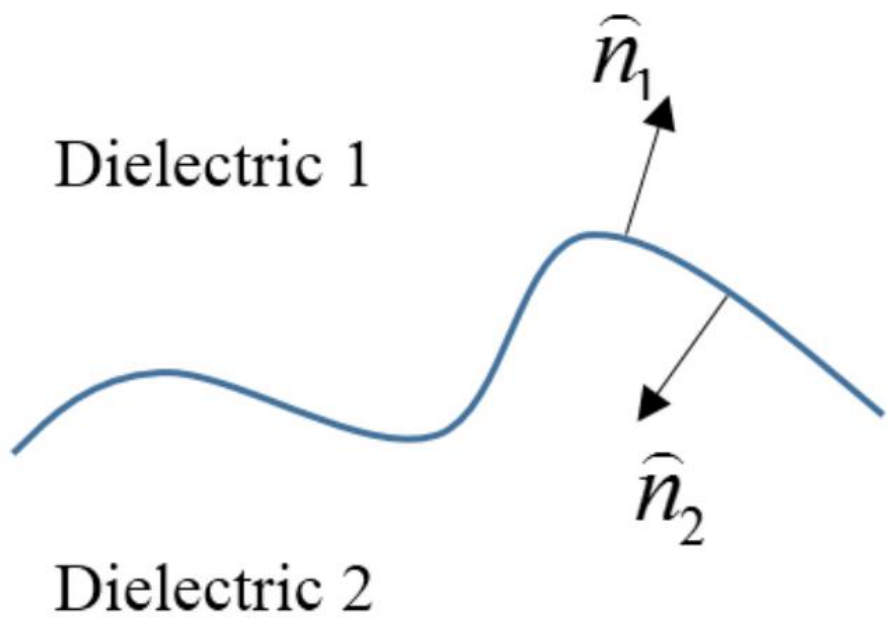

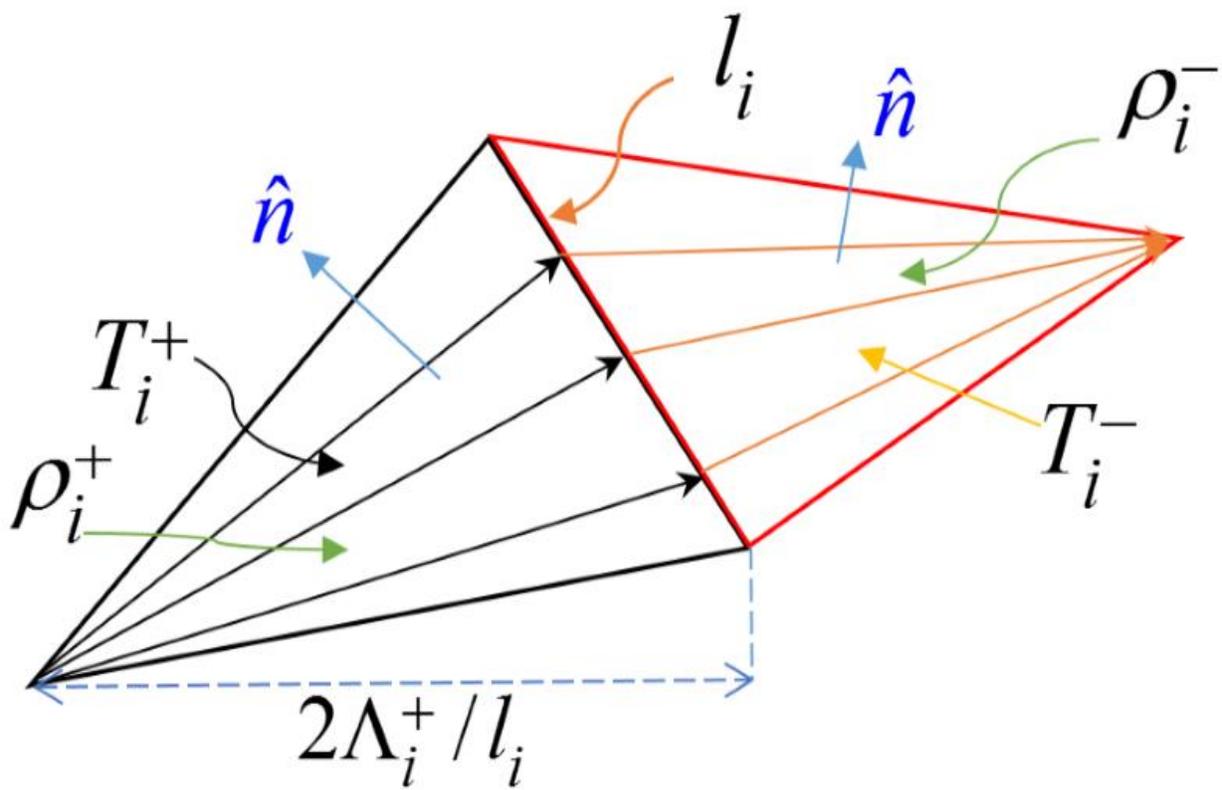

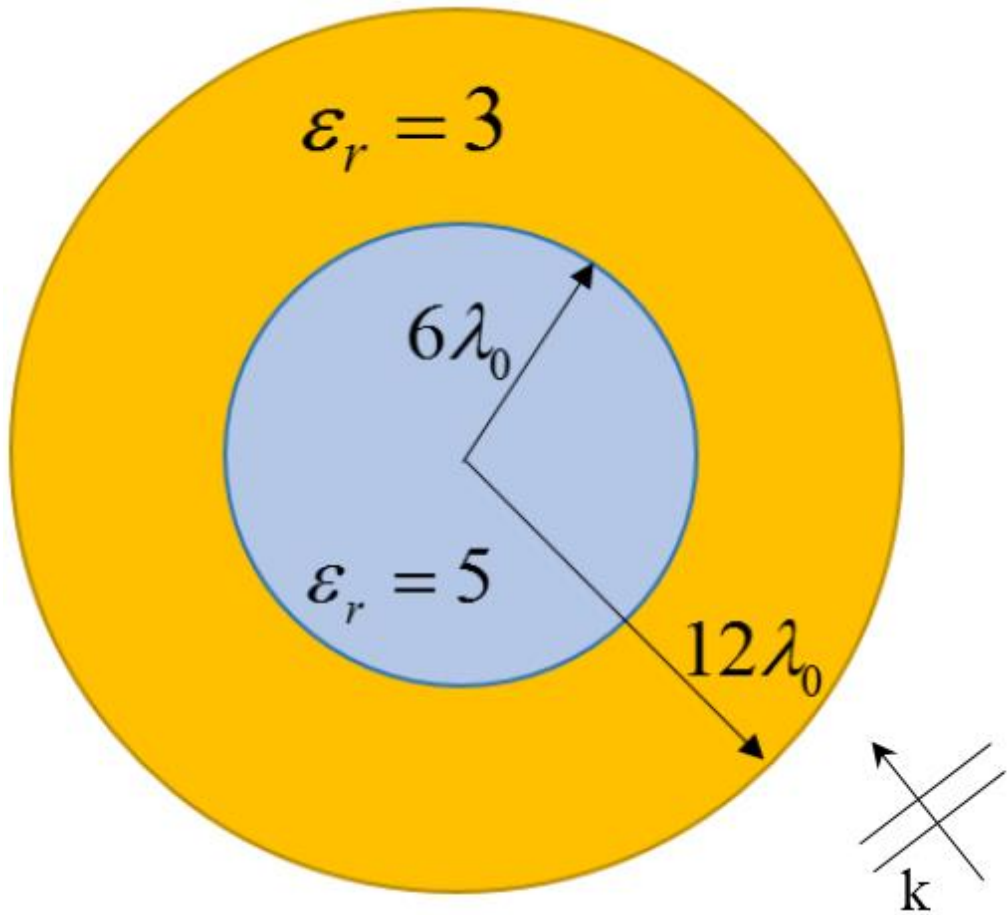
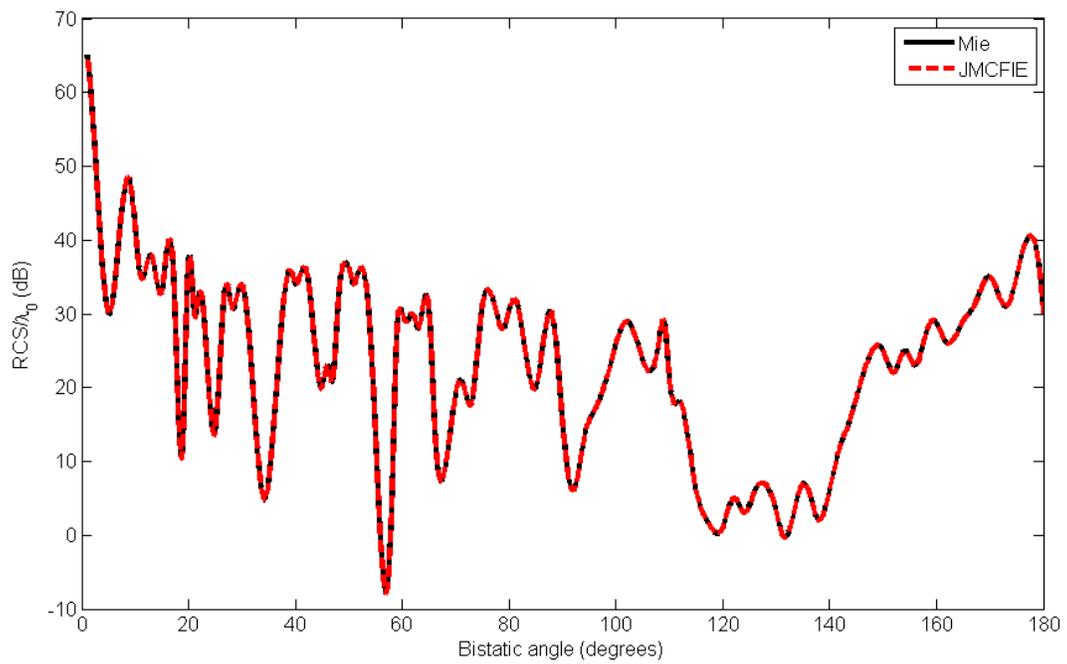

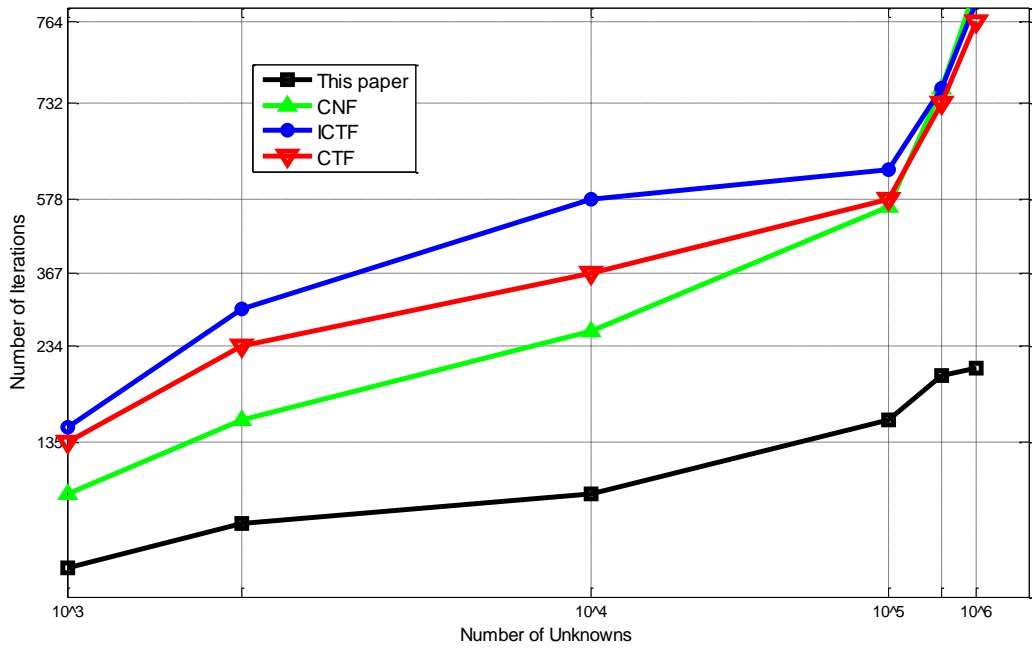

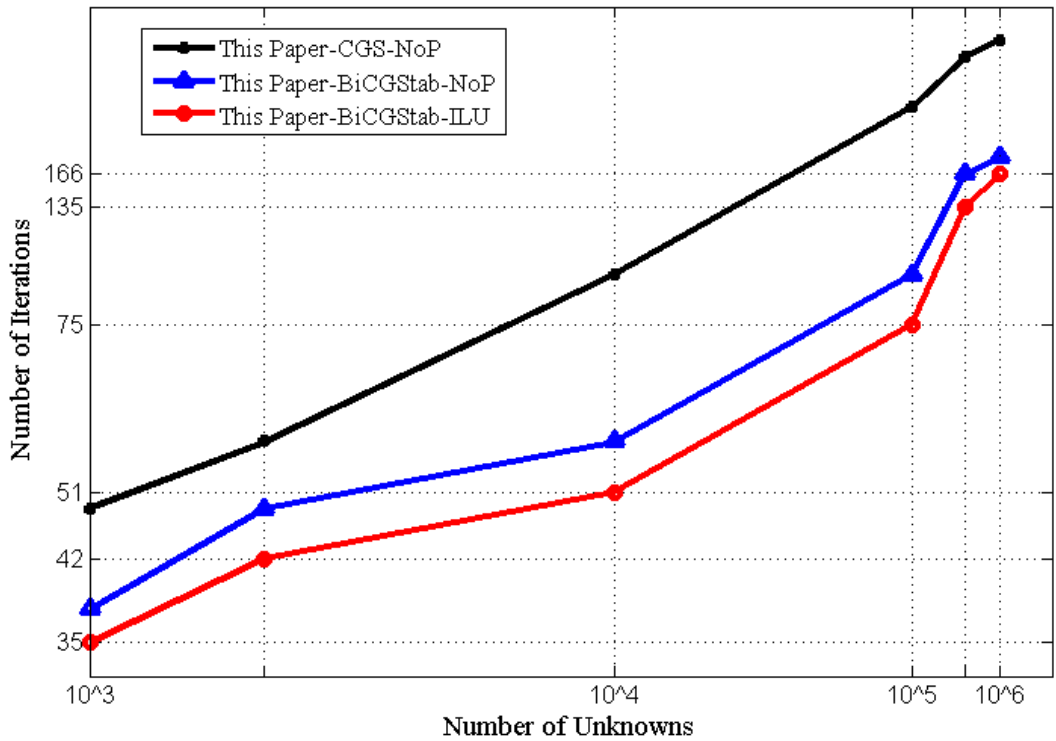

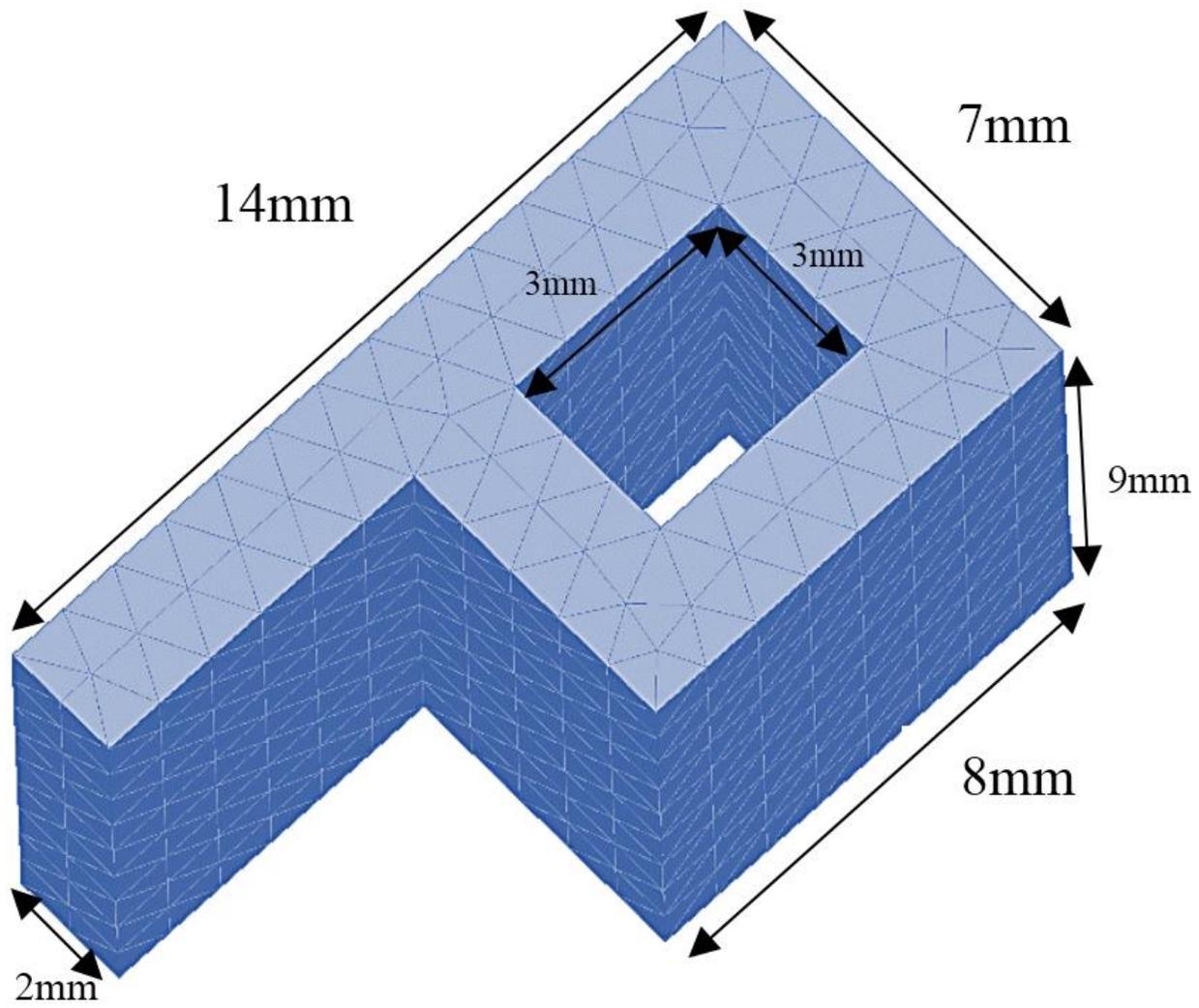

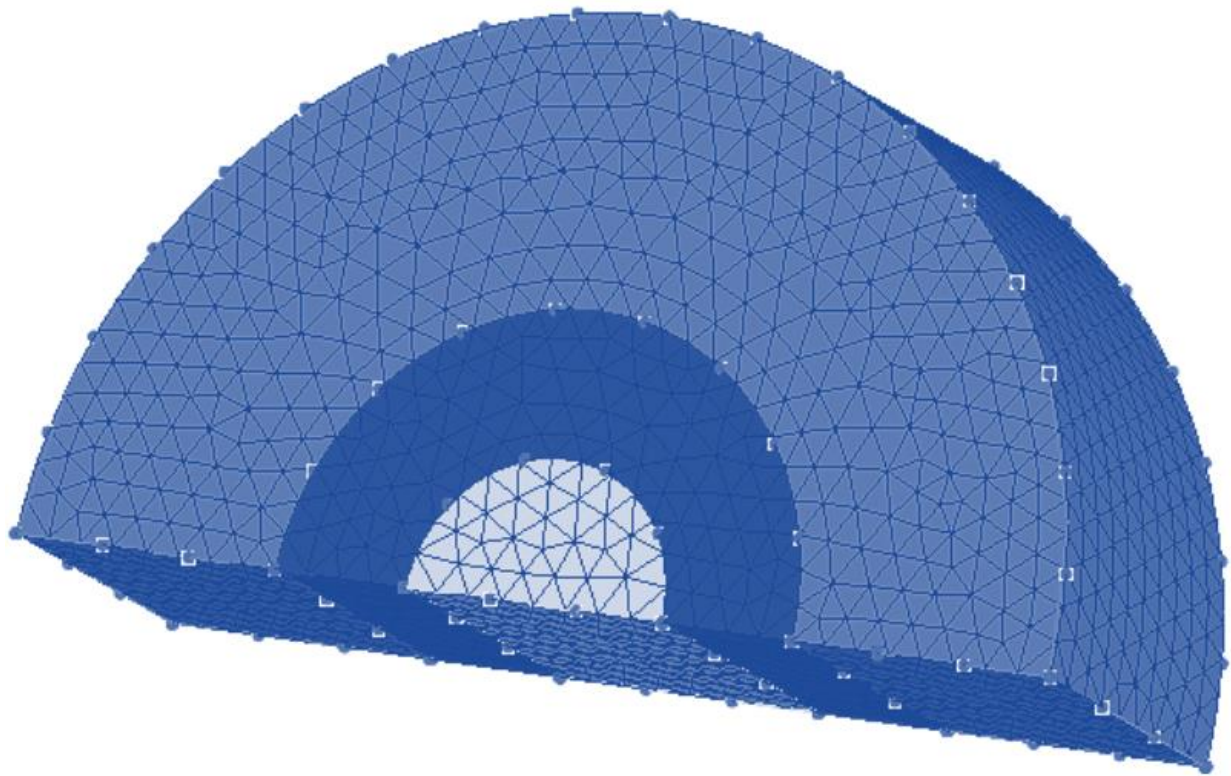

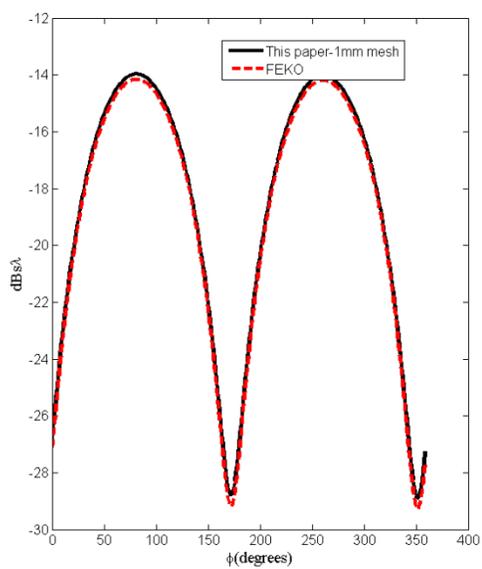 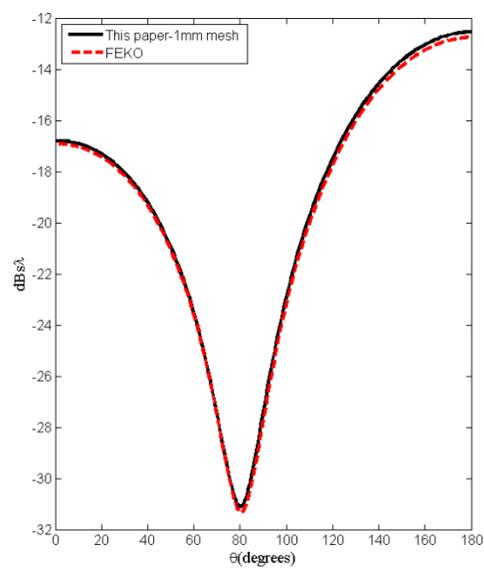

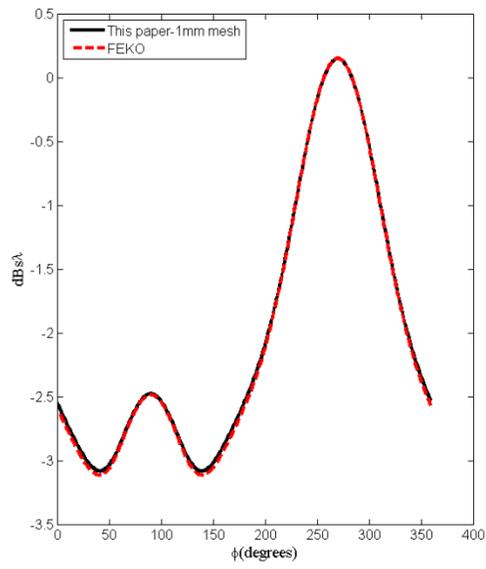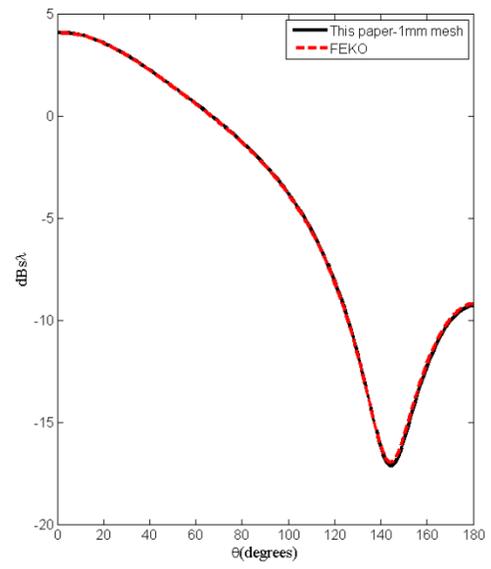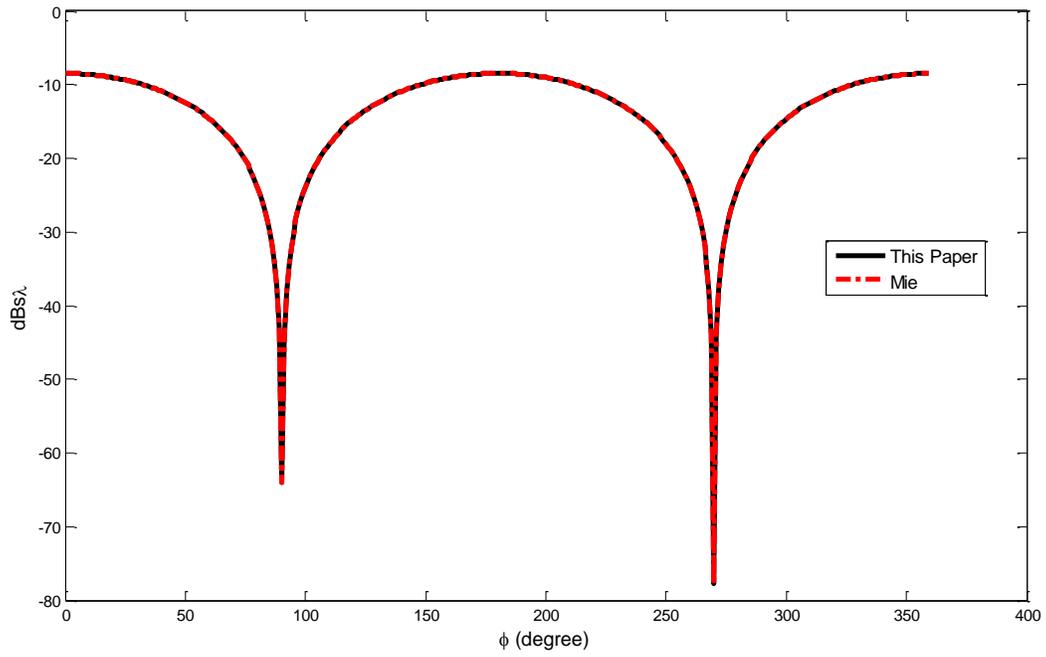